\documentclass[%
preprint,
 amsmath,amssymb,
 aps,
]{revtex4-1}

\usepackage{graphicx}
\usepackage{dcolumn}
\usepackage{bm}

\begin{document}


\title{The optimal positive operator-valued measure for state discrimination}

\author{Wei Li$^{1,2,3}$}
\author{Shengmei Zhao$^{1,2}$}%
 \email{zhaosm@njupt.edu.cn}
\affiliation{$^{1}$Nanjing University of Posts and Telecommunications, Institute of Signal Processing and Transmission, Nanjing, 210003, China.}%
\affiliation{$^{2}$Nanjing University of Posts and Telecommunications, Key Lab Broadband Wireless Communication and Sensor Network, Ministy of Education, Nanjing, 210003, China.}%
\affiliation{$^{3}$National Laboratory of Solid State Microstructures, Nanjing University, Nanjing 210093, China.}%


\date{\today}

\begin{abstract}

Evaluating the amount of information obtained from non-orthogonal quantum states is an important topic in the field of quantum information. The commonly used evaluation method is Holevo bound, which only provides a loose upper bound for quantum measurement. In this paper, we provide a theoretical study of the positive operator-valued measure (POVM) for discriminating nonorthogonal states. We construct a generalized POVM measurement operation, and derive the optimal one for state discrimination by Lagrange multiplier method. With simulation, we find that the optimal POVM measurement provides a tight upper bound for state discrimination, which is significantly lower than that predicted by Holevo bound. The derivation of optimal POVM measurement will play an important role in the security research of quantum key distribution.
\end{abstract}

\pacs{Valid PACS appear here}
\maketitle

\section{Introduction}
\par Quantum measurement is a way to obtain information from unknown quantum systems, which is composed of a set of measurement operators that sum to the identity operator\cite{nielsen2002quantum}. The simplest measurement operator is the projection operator, which is composed of a set of complete orthogonal bases, whose eigenvalue can be continuous or discrete, depending on the specific measured system. In quantum measurement, we are usually only interested in the state of the system, so we only pay attention to the probability of the result obtained by each measurement. A generalized quantum measurement is positive operator-valued measure (POVM)\cite{brandt1999positive,hamieh2004positive,winter2004extrinsic,ziman2008process}, in which the measurement operators can be non-orthogonal and completely described with the help of an ancillary state. 

\par Quantum measurement has attracted extensive attentions in the field of quantum information. A typical example is evaluating the quantum channel attack in quantum key distribution (QKD) technology, where the eavesdropper carries out the channel attacks by interacting his ancillary states with the transmitted states\cite{scarani2009security,xu2020secure}. After the interaction, the eavesdroppers select appropriate quantum measurements for the stored ancillary quantum states to eavesdrop on as much information as possible about the transmitted quantum states. In quantum channel attacks, the ancillary quantum states stored by the eavesdropper are usually non-orthogonal. There are three main types of quantum channel attacks. The first is the coherent attack, which is a conceptual ideal high-dimensional channel attack scheme\cite{lo1999unconditional,shor2000simple,sheridan2010finite,lo2012measurement,ma2018phase}. The key rate for this attack is provided by the entanglement purification protocol (EPP), which uses the quantum error correction code to purify the virtual Bell states, but EPP does not involve any quantum measurement. The second is the collective attack scheme, which is an asymptotically optimal quantum channel attack\cite{biham1997security,biham2002security,boyer2009security,acin2007device,pironio2009device,li2019phase}. The eavesdropping capability of the eavesdropper is characterized by Holevo bound\cite{holevo1973bounds,shirokov2019upper,roga2010universal}. However, this method does not involve any specific measurement operation, so only an loose upper bound of the eavesdropping is provided. The third is the individual attack scheme\cite{fuchs1997optimal,griffiths1997optimal,bechmann1999incoherent}, in which the eavesdropper's attack capability is characterized by projection measurement. However, in individual attack, there is no detailed discussion on the generalized quantum measurement, so it is still uncertain whether the quantum measurement in this attack scheme is optimal. Therefore, finding the optimal quantum measurement scheme to characterize the tight upper bound of channel attack is an urgent problem to evaluate the capability of Eve's channel attacks.

\par In this paper, we give a theoretical study on the optimal POVM measurement for discriminating two non-orthogonal quantum states. Firstly, we construct a generalized POVM measurement, which is composed of bipartite interaction with an ancillary quantum state and followed a projection measurement of measured states after that interaction. Here, POVM measurement is equivalent to a classical channel, and its measurement capability is expressed as the mutual information. Next, we use Lagrange multiplier method to study the relationship between the dimension of the optimal projection measurement and the number of measured states, as well as find out the best POVM measurement. Finally, we compared the optimal POVM measurement with the Holevo bound for the quantum measurement of two non-orthogonal states.

\section{Generalized POVM measurement for non-orthogonal states}

\par Imagine a game in which Alice tosses a coin, and send states $\left | p \right \rangle$ or $\left | q \right \rangle$ to Eve depending on whether the coin is head or tail. States $\left | p \right \rangle$ and $\left | q \right \rangle$ are non-orthogonal, and their inner product is $\left \langle p | q \right \rangle=\cos\alpha$, where $\alpha$ is the angle between states $\left | p \right \rangle$ and $\left | q \right \rangle$. Let's further assume that the coin was tampered, the probability of occurrence of the head is $p$, the probability of occurrence of the tail is $q$, and they satisfy the normalization condition $p+q=1$. So the mixed state that Eve received is
\begin{equation}
\rho = p \rho_{p} + q \rho_{q},
\end{equation}
where $\rho_{p}=\left | p \right \rangle \left \langle p \right |$ and $\rho_{q}=\left | q \right \rangle \left \langle q \right |$. Eve's task is to find an appropriate POVM measurement, a generalized quantum measurement scheme, to measure the states$\left | p \right \rangle$ and $\left | q \right \rangle$, and infer whether Alice's coin is head or tail according to the measurement results. So, what is the optimal POVM measurement to maximize the accuracy of Eve's speculation?

\par A general POVM measurement can be formulated as a bipartite interaction with an ancillary state initiated as $\left | 0 \right \rangle_{A}$\cite{nielsen2002quantum},
\begin{equation}
\begin{split}
U \left | p \right \rangle \left | 0 \right \rangle_{A} =& \sum_{i}^{n} M_{i} \left | p \right \rangle \left | i \right \rangle_{A}=\sum_{i}^{n} \sqrt{p_{i}} \left | p_{i} \right \rangle \left | i \right \rangle_{A},\\
U \left | q \right \rangle \left | 0 \right \rangle_{A} =& \sum_{i}^{n} M_{i} \left | q \right \rangle \left | i \right \rangle_{A}=\sum_{i}^{n} \sqrt{q_{i}} \left | q_{i} \right \rangle \left | i \right \rangle_{A},
\end{split}
\end{equation}
where $\left | i \right \rangle_{A}$ is a set of orthogonal bases with $\left \langle i | j \right \rangle_{A}=0$, $M_{i}$ is a set of unitary operators that satisfies $\sum_{i} M^{\dagger}_{i}M_{i}=I$ with $I$ an identity operator, $p_{i}$ and $q_{i}$ are the probabilities of the occurence of state $\left | i \right \rangle_{A}$ for measurement on $\left | p \right \rangle$ and $\left | q \right \rangle$ with $p_{i}=\left \langle p \right | M^{\dagger}_{i}M_{i}\left | p \right \rangle$, $q_{i}=\left \langle q \right | M^{\dagger}_{i}M_{i}\left | q \right \rangle$. Once state $\left | i \right \rangle_{A}$ is measured, states $\left | p_{i} \right \rangle$ and $\left | q_{i} \right \rangle$ should be distinguished to obtain the information about $\left | p \right \rangle$ and $\left | q \right \rangle$. This step is finished by a projection measurement by choosing a set of bases $\left | j \right \rangle$ with $j\in \{ 0,1,2,\cdots \}$. Then the total quantum measurement for states $\left | p \right \rangle$ and $\left | q \right \rangle$ can be formulated as a conditional probability matrix
\begin{equation}
p_{M}=
\begin{bmatrix}
p_{1}\left | a_{1,0} \right |^{2} & p_{1}\left | a_{1,1} \right |^{2} & \cdots,  & p_{2}\left | a_{2,0} \right |^{2} & p_{2}\left |a_{2,1}\right |^{2} & \cdots, & \cdots,  & p_{n}\left | a_{n,0}\right |^{2} & p_{n}\left | a_{n,1} \right |^{2} & \cdots\\ 
q_{1}\left | b_{1,0}\right |^{2} & q_{1}\left | b_{1,1} \right |^{2} & \cdots,  & q_{2}\left | b_{2,0} \right |^{2} & q_{2}\left |b_{2,1} \right |^{2} & \cdots, & \cdots,  & q_{n}\left | b_{n,0} \right |^{2} & q_{n}\left | b_{n,1} \right |^{2} & \cdots
\end{bmatrix},
\end{equation}
where $a_{i,j}=\left \langle j | p_{i} \right \rangle$ and $b_{i,j}=\left \langle j | q_{i} \right \rangle$, $\sum_{j}a_{i,j}^{*}b_{i,j}=\left \langle p_{i} | q_{i} \right \rangle=\cos \alpha_{i}$ with $\alpha_{i}$ the angle between states $\left | p_{i} \right \rangle$ and $\left | q_{i} \right \rangle$, the measurement matrix of the $k$-th sub-channel is
\begin{equation}
p_{M_{k}}=
\begin{bmatrix}
\left | a_{k,0} \right |^{2} & \left | a_{k,1} \right |^{2} & \cdots\\ 
\left | b_{k,0}\right |^{2} & \left | b_{k,1} \right |^{2} & \cdots
\end{bmatrix}.
\end{equation}
As the phase factors in $\left | p \right \rangle$ and $\left | q \right \rangle$ have no effect on the measurement matrix and the module of the inner product between them, we can assume that both of their phases are 0, and the matrix elements $a_{i,j}$ and $b_{i,j}$ can be viewed as real numbers. Therefore, the POVM measurement matrix can be regarded as a channel, and the measurement scheme that can achieve the channel capacity is optimal.

\section{Dimension of optimal projection measurement}

\par Before deriving the channel capacity of the measurement matrix, we first study the relationship between the dimension of projection measurement and the number of measured states. We take the $i$-th subchannel as an example, where $\dfrac{pp_{i}}{pp_{i}+qq_{i}}$ and $\dfrac{qq_{i}}{pp_{i}+qq_{i}}$ are the normalized input probabilities of the channel, the mutual information for the $i$-th measurement matrix is
\begin{equation}
\chi_{i}=\sum_{j}\left ( pp_{i}a_{i,j}^{2} \log_{2} \frac{\left ( pp_{i}+qq_{i} \right)a_{i,j}^{2}}{pp_{i}a_{i,j}^{2}+qq_{i}b_{i,j}^{2}}+qq_{i}b_{i,j}^{2} \log_{2} \frac{\left ( pp_{i}+qq_{i} \right)b_{i,j}^{2}}{pp_{i}a_{i,j}^{2}+qq_{i}b_{i,j}^{2}} \right ),
\end{equation}
which subject to conditions $\sum_{j}a_{i,j}b_{i,j}=\cos \alpha_{i}$ and $\sum_{j} \left ( a_{i,j}^{2}+b_{i,j}^{2} \right )=2$. The maximum value of $\chi_{i}$ can be calculated with Lagrange multiplier method. Here we construct the Lagrange function for the $i$-th subchannel
\begin{equation}
L_{i}=\chi_{i}+\lambda_{1}\left ( \cos \alpha_{i}-\sum_{j}a_{i,j}b_{i,j} \right )+\lambda_{2}\left ( 2-\sum_{j} a_{i,j}^{2}+\sum_{j} b_{i,j}^{2} \right ),
\end{equation}
where $\lambda_{1}$ and $\lambda_{2}$ are two constants. Around the maximum value of $\chi_{i}$, $L_{i}$ satisfies
\begin{equation}
\begin{split}
\frac{\partial L_{i}}{\partial a_{i,j}}=& 2pp_{i}a_{i,j} \log_{2}\frac{\left ( pp_{i}+qq_{i} \right)a_{i,j}^{2}}{pp_{i}a_{i,j}^{2}+qq_{i}b_{i,j}^{2}}-\lambda_{1} b_{i,j}-2\lambda_{2}a_{i,j},\\
\frac{\partial L_{i}}{\partial b_{i,j}}=& 2qq_{i}b_{i,j} \log_{2}\frac{\left ( pp_{i}+qq_{i} \right)b_{i,j}^{2}}{pp_{i}a_{i,j}^{2}+qq_{i}b_{i,j}^{2}}-\lambda_{1} a_{i,j}-2\lambda_{2}b_{i,j}.
\end{split}
\end{equation}
By summing the equations in Eq. (7), we have
\begin{equation}
\begin{split}
2\log_{2}\left (\frac{pp_{i}a_{i,j}}{b_{i,j}}+\frac{qq_{i}b_{i,j}}{a_{i,j}} \right )+\frac{\lambda_{1}}{2pqp_{i}q_{i}}\left (\frac{pp_{i}a_{i,j}}{b_{i,j}}+\frac{qq_{i}b_{i,j}}{a_{i,j}} \right )&\\
+\frac{\lambda_{2}\left ( pp_{i}+qq_{i} \right )}{pqp_{i}q_{i}}+2\log_{2}\left ( pp_{i}+qq_{i} \right )&=0.
\end{split}
\end{equation}
From Eq. (8), we can see that $\dfrac{pp_{i}a_{i,j}}{b_{i,j}}+\dfrac{qq_{i}b_{i,j}}{a_{i,j}}$ is a function of $pp_{i}qq_{i}$ and $pp_{i}+qq_{i}$,
\begin{equation}
\frac{pp_{i}a_{i,j}}{b_{i,j}}+\frac{qq_{i}b_{i,j}}{a_{i,j}}=f\left ( pp_{i}qq_{i}, pp_{i}+qq_{i} \right ).
\end{equation}
Once the values of $pp_{i}$ and $qq_{i}$ are provided, there are only two solutions for Eq. (9) that $\dfrac{a_{i,j}}{b_{i,j}}=\{ g_{1} \left ( pp_{i}qq_{i}, pp_{i}+qq_{i} \right ), g_{2} \left ( pp_{i}qq_{i}, pp_{i}+qq_{i} \right ) \}$, whose value determines the placement of the projection bases with respect to $\left | p_{i} \right \rangle$ and $\left | q_{i} \right \rangle$.

\par Let's set the plane formed by states $\left | p_{i} \right \rangle$ and $\left | q_{i} \right \rangle$ as $\Gamma_{i}$, any projection basis $\left | j \right \rangle$ can be decomposed into a component $\left | j \right \rangle_{\parallel}$ parallel to $\Gamma_{i}$ and a component $\left | j \right \rangle_{\perp}$ perpendicular to $\Gamma_{i}$, i.e. $\left | j \right \rangle = c_{j} \left | j \right \rangle_{\parallel}+d_{j} \left | j \right \rangle_{\perp}$. Then we can get $a_{i,j}=c_{j}\left \langle p_{i} | j \right \rangle_{\parallel}$, $b_{i,j}=c_{j}\left \langle q_{i} | j \right \rangle_{\parallel}$ and $\dfrac{a_{i,j}}{b_{i,j}}=\dfrac{\left \langle p_{i} | j \right \rangle_{\parallel}}{\left \langle q_{i} | j \right \rangle_{\parallel}}$. If we choose the first two bases $\left | 0 \right \rangle$ and $\left | 1 \right \rangle$ that satisfy $\dfrac{\left \langle p_{i} | 0 \right \rangle_{\parallel}}{\left \langle q_{i} | 0 \right \rangle_{\parallel}}=g_{1} \left ( pp_{i}qq_{i},pp_{i}+qq_{i} \right )$ and $\dfrac{\left \langle p_{i} | 1 \right \rangle_{\parallel}}{\left \langle q_{i} | 1 \right \rangle_{\parallel}}=g_{2} \left ( pp_{i}qq_{i},pp_{i}+qq_{i} \right )$, then a third basis $\left | 2 \right \rangle$ that satisfies the same condition must parallel to $\left | 0 \right \rangle_{\parallel}$ or $\left | 1 \right \rangle_{\parallel}$ within $\Gamma_{i}$, which will not guarantee the orthogonality of the bases. So a third basis does not exist, the dimension of optimal projection measurement is equal to the number of measured states.

\section{Derivation of optimal POVM measurement}
With the conclusion of section III, the measurement matrix of Eq. (3) can be written as
\begin{equation}
p_{M}=
\begin{bmatrix}
p_{1} a_{1,0}^{2} & p_{1} a_{1,1}^{2} ,  & p_{2} a_{2,0}^{2} & p_{2}a_{2,1}^{2} , & \cdots,  & p_{n} a_{n,0}^{2} & p_{n}a_{n,1}^{2} \\ 
q_{1} b_{1,0}^{2} & q_{1} b_{1,1}^{2} ,  & q_{2} b_{2,0}^{2} & q_{2}b_{2,1}^{2} , & \cdots,  & q_{n} b_{n,0}^{2} & q_{n} b_{n,1}^{2}
\end{bmatrix},
\end{equation}
whose input probabilities are $p$ and $q$. Then the mutual information for the total POVM measurement matrix is
\begin{equation}
\chi=\sum{i,j}\left [ pp_{i}a_{i,j}^{2}\log_{2} \frac{p_{i}a_{i,j}^{2}}{pp_{i}a_{i,j}^{2}+qq_{i}b_{i,j}^{2}}+ qq_{i}b_{i,j}^{2}\log_{2} \frac{q_{i}b_{i,j}^{2}}{pp_{i}a_{i,j}^{2}+qq_{i}b_{i,j}^{2}} \right ],
\end{equation}
which subject to conditions $\sum_{i}\sqrt{p_{i}q_{i}}a_{i,j}b_{i,j}=\cos\alpha$ and $\sum_{i,j}\left ( pp_{i}a_{i,j}^{2}+qq_{i}b_{i,j}^{2} \right )=1$, where $i\in \{ 1,2,\cdots,n\}$ and $j\in \{ 0,1\}$. Here again, we use Lagrange multiplier method to derive the maximum value of $\chi$. The Lagrange function for $\chi$ is
\begin{equation}
L = \chi + \lambda_{1}\left ( \cos \alpha-\sum_{i}\sqrt{p_{i}q_{i}}a_{i,j}b_{i,j} \right )+\lambda_{2}\left ( 1-\sum_{i,j}\left ( pp_{i}a_{i,j}^{2}+qq_{i}b_{i,j}^{2} \right ) \right ).
\end{equation}
Around the maximum value of $\chi$, we have
\begin{equation}
\begin{split}
\frac{\partial L}{\partial a_{i,j}}=& 2pp_{i}a_{i,j} \log_{2}\frac{p_{i}a_{i,j}^{2}}{pp_{i}a_{i,j}^{2}+qq_{i}b_{i,j}^{2}}-\lambda_{1} pq\sqrt{p_{i}q_{i}}b_{i,j}-\lambda_{2}2pp_{i}a_{i,j},\\
\frac{\partial L}{\partial b_{i,j}}=& 2qq_{i}b_{i,j} \log_{2}\frac{q_{i}b_{i,j}^{2}}{pp_{i}a_{i,j}^{2}+qq_{i}b_{i,j}^{2}}-\lambda_{1} pq\sqrt{p_{i}q_{i}}a_{i,j}-\lambda_{2}2qq_{i}b_{i,j}.
\end{split}
\end{equation}
By summing the equations in Eq. (13), we get
\begin{equation}
\log_{2}\left ( p\sqrt{\frac{p_{i}}{q_{i}}}\frac{a_{i,j}}{b_{i,j}}+q\sqrt{\frac{q_{i}}{p_{i}}}\frac{b_{i,j}}{a_{i,j}} \right)+\frac{\lambda_{1}}{4}\left ( p\sqrt{\frac{p_{i}}{q_{i}}}\frac{a_{i,j}}{b_{i,j}}+q\sqrt{\frac{q_{i}}{p_{i}}}\frac{b_{i,j}}{a_{i,j}} \right)+\lambda_{2}=0.
\end{equation}
From Eq. (14), we can see that $p\sqrt{\dfrac{p_{i}}{q_{i}}}\dfrac{a_{i,j}}{b_{i,j}}+q\sqrt{\dfrac{q_{i}}{p_{i}}}\dfrac{b_{i,j}}{a_{i,j}}$ is a constant. Let's set $p\sqrt{\dfrac{p_{i}}{q_{i}}}\dfrac{a_{i,j}}{b_{i,j}}+q\sqrt{\dfrac{q_{i}}{p_{i}}}\dfrac{b_{i,j}}{a_{i,j}}=\gamma$, we have the following equality
\begin{equation}
\frac{pp_{i}a_{i,j}}{b_{i,j}}+\frac{qq_{i}b_{i,j}}{a_{i,j}}=\frac{pp_{i}a_{i,j}^{2}+qq_{i}b_{i,j}^{2}}{a_{i,j}b_{i,j}}=\gamma \sqrt{p_{i}q_{i}},
\end{equation}
here $p$ and $q$ are regarded as known constants, $p_{i}$ and $q_{i}$ are variables that depend on the subchannel index $i$. From Eq. (15), we further get
\begin{equation}
\sum_{j}\left ( pp_{i}a_{i,j}^{2}+qq_{i}b_{i,j}^{2} \right )=pp_{i}+qq_{i}=\sum_{j} \gamma \sqrt{p_{i}q_{i}}a_{i,j}b_{i,j} =\gamma \sqrt{p_{i}q_{i}} \cos \alpha_{i}.
\end{equation}

\par Because the total mutual information of the POVM measurement is the sum of the mutual information of each subchannel, and the subchannels are independent of each other, when $\chi$ reaches its maximum, $\chi_{i}$ for the $i$-th subchannel should reach its maximum as well. Combine Eqs. (9) and (15), we have $pp_{i}+qq_{i}=h\left (\sqrt{p_{i}q_{i}} \right)$. It's easy to see that polynomials $pp_{i}+qq_{i}$ and $\sqrt{p_{i}q_{i}}$ are in the same order, then $pp_{i}+qq_{i}\propto \sqrt{p_{i}q_{i}}$, so we can conclude that $p\sqrt{\dfrac{p_{i}}{q_{i}}}+q\sqrt{\dfrac{q_{i}}{p_{i}}}$ is a constant,
\begin{equation}
p\sqrt{\frac{p_{i}}{q_{i}}}+q\sqrt{\frac{q_{i}}{p_{i}}}=\gamma_{1}=\gamma \cos \alpha_{i}.
\end{equation}
It is easy to see that the value of $\cos \alpha_{i}$ is independent of the subchannel index $i$. Here we set $\cos \alpha_{i}=\gamma_{2}$, then the equality $\sum_{i}\sqrt{p_{i}q_{i}}\cos \alpha_{i}=\gamma_{2} \sum_{i}\sqrt{p_{i}q_{i}}=\cos\alpha$ can be obtained. Because $\sum_{i}\sqrt{p_{i}q_{i}}\leq 1$, so $\gamma_{2} = \dfrac{\cos \alpha}{\sum_{i}\sqrt{p_{i}q_{i}}} \geq \cos \alpha$, where the equality holds when $p_{i}=q_{i}$. As the discrimination of states $\left | p_{i} \right \rangle$ and $\left | q_{i} \right \rangle$ depends on the angle between them, the smaller $\gamma_{2}$ is, the larger $\alpha_{i}$ we have, the more we can distinguish $\left | p_{i} \right \rangle$ and $\left | q_{i} \right \rangle$. So the optimal POVM measurement is that for each subchannel, $p_{i}=q_{i}$, $\left \langle p_{i} | q_{i} \right \rangle=\cos \alpha_{i} =\cos \alpha$ and $p\dfrac{a_{i,j}}{b_{i,j}}+q\dfrac{b_{i,j}}{a_{i,j}}$ is independent of the subchannel index $i$. In this case, the optimal POVM measurement is equivalent to a projection measurement on the initial states. Therefore, in this game, Eve's optimal POVM measurement strategy is to use the projection measurement that meets the above conditions, which could maximize the accuracy of her guess about the result of Alice's coin toss.

\begin{figure}[ht]
\centering
\includegraphics[width=100mm]{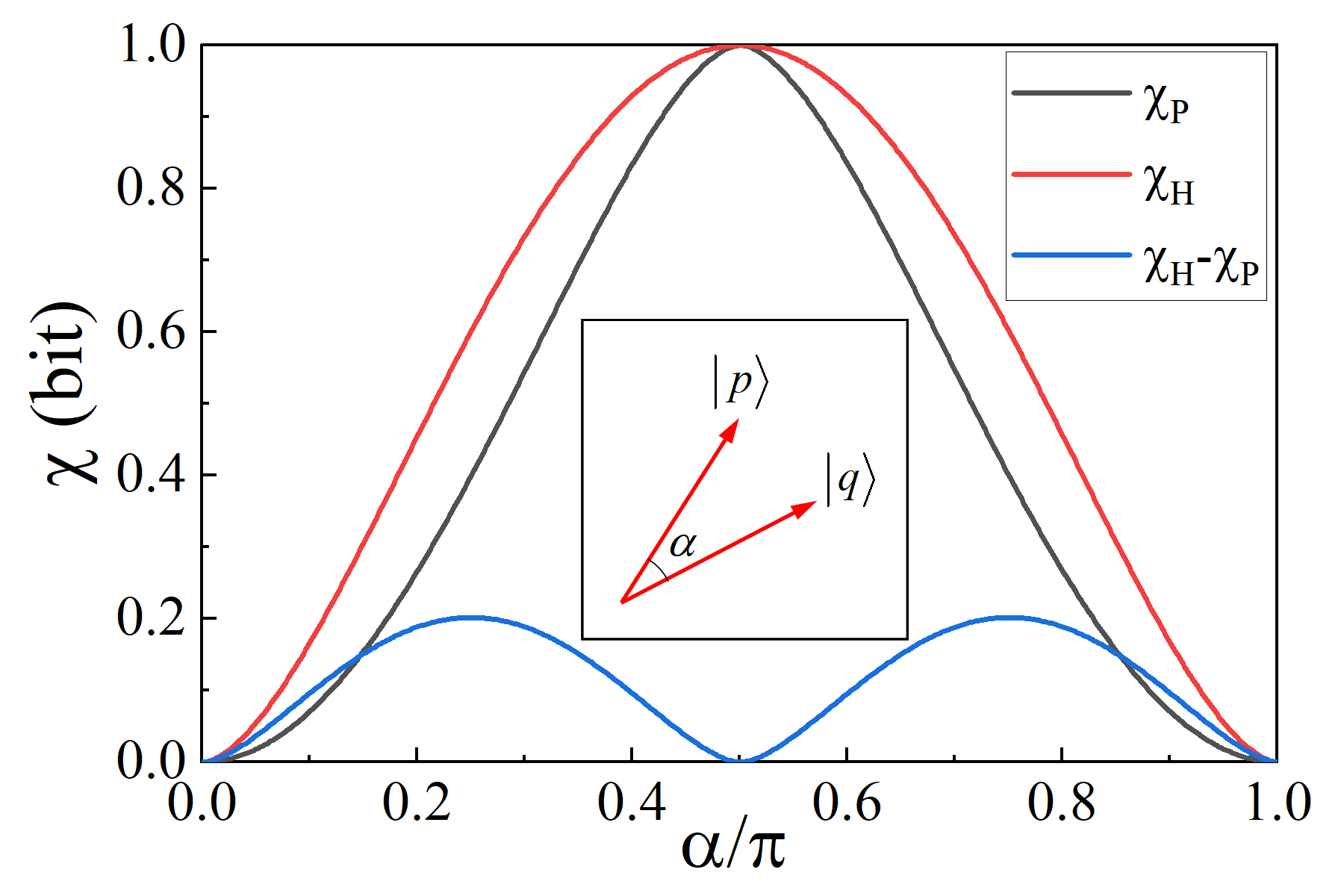}
\caption{Comparison of the amount of information evaluated by optimal POVM measurement $\chi_{P}$ and Holevo bound $\chi_{H}$ as well as the difference between them $\chi_{H}-\chi_{P}$ from two non-orthogonal states with respect to the angle $\alpha$ between them.}
\label{Fig. 1}
\end{figure}

\par Another more practical role played by POVM measurement is to evaluate the eavesdropping capability of quantum channel attacks by distinguishing non-orthogonal states in the security analysis of QKD. Next, we compare POVM measurement and Holevo bound, which are two commonly used information quantization methods in the field of quantum information. For two equiprobability mixed states $\left | p \right \rangle$ and $\left | q \right \rangle$, the maximum amount of information that can be obtained by POVM measurement is $\chi_{P}=1-H\left ( \dfrac{1+\sin \alpha}{2} \right )$, while the maximum amount of information that can be obtained by Holevo bound is $\chi_{H}=H\left ( \dfrac{1+\cos \alpha}{2} \right )$, where $H\left (x \right )$ is Shannon entropy with $H\left (x \right )=-x\log_{2}x-\left (1-x \right )\log_{2}\left (1-x\right)$ and $\left \langle p | q \right \rangle=\cos \alpha$. Fig .1 shows the comparison of $\chi_{P}$ and $\chi_{H}$ with respect to $\alpha$, where Holevo bound is believed to provide an upper bound for information extraction\cite{holevo1973bounds}. The black solid line is the amount of information evaluated by the optimal POVM measurement the red solid line is that evaluated by Holevo bound, and the blue solid line is the difference between them. From this figure, we can see that the information evaluated by Holevo bound is always larger than that evaluated by POVM measurement except for $\alpha=0$ or $\alpha=\dfrac{\pi}{2}$, where these two states can not be distinguished or can be completely distinguished. Therefore, the Holevo bound provides a loose upper bound of eavesdropping capability of quantum channel attacks, while the optimal POVM measurement provides a tight bound of that. With the optimal POVM measurement derived above, a higher key rate can be obtained compared with the methods based on EPP and Holevo bound.

\section{Conclusion}
In this paper, We derive the optimal POVM measurement scheme for two arbitrary non-orthogonal states. Here we find that the measurement matrix of POVM measurement can be equivalent to a classical channel, projection measurements one of the optimal POVM measurement, and the dimension of projection measurement is equal to the number of states to be measured. Compared with Holevo bound, the optimal POVM measurement provides a tight bound on the amount of information obtained from non-orthogonal states. Based on this, the optimal POVM measurement will play an important role in improving the key rate of practical QKD.
\section*{Acknowledgments}
This work is supported by China Postdoctoral special funding project (2020T130289), the National Natural Science Foundation of China (No. 61871234).

\bibliography{reference}

\end{document}